\documentstyle[aps,preprint,tighten]{revtex}
\def\be{\begin{equation}}
\def\ee{\end{equation}}

\input{boxedeps.tex}
\SetRokickiEPSFSpecial
\HideDisplacementBoxes

\begin{document}
\draft
\def\Li{$^{11}$Li}
\def\He6{$^6$He}
\def\C12{$^{12}$C}
\title{Nuclear Breakup of Borromean Nuclei}
\author{G. F. Bertsch, and K. Hencken}
\address{Institute for Nuclear Theory, University of Washington,
Seattle, Washington 98195}
\author{and H. Esbensen}
\address{Physics Division, Argonne National Laboratory, Argonne, 
Illinois 60439}
\maketitle
 
\begin{abstract} We study the eikonal model for the nuclear-induced
breakup of Borromean nuclei, using \Li{} and \He6{} as examples.  The
full eikonal model is difficult to realize because of six-dimensional
integrals, but a number of simplifying approximations are found to be
accurate. The integrated diffractive and one-nucleon stripping cross
sections are rather insensitive to the neutron--neutron correlation,
but the two-nucleon stripping does show some dependence on the
correlation.  The distribution of excitation energy in the
neutron--core final state in one-neutron stripping reactions is quite
sensitive to the shell structure of the halo wave function.
Experimental data favor models with comparable amounts of $s$- and
$p$-wave in the \Li{} halo. \end{abstract}

\section{Introduction}

Halo nuclei having a very weakly bound neutron pair (often referred to
as Borromean nuclei) are interesting objects but they are difficult to
study experimentally.  Secondary interactions in radioactive beams
have been an important tool, with Coulomb excitation providing
quantitative data about the excitation properties\cite{sa93,sh95}.
Nuclear excitation is also important from an experimental point of
view, but the theoretical interpretation of nuclear reaction cross
sections deserves closer attention.  In this work we attempt to make a
link, as quantitative as possible, between the nuclear excitation
observables and the fundamental properties of a Borromean nucleus.
The fact that correlations can play an important role makes this goal
more difficult than for a nucleus with a single-nucleon halo.  On the
experimental side, we have been inspired by the work on \Li{} carried
out at Ganil, NSCL, RIKEN and most recently at GSI. The extremely large
Coulomb breakup cross section shows the halo character of the nucleus,
but the details of its wave function have been controversial.
Starting from the shell model, two of us
\cite{be91} constructed a wave function that fit many Coulomb
excitation measurements \cite{es93}.  It had a dominant $p_{1/2}^2$
shell configuration, as one expects from Hartree-Fock theory.
However, several measurements (see for example ref. \cite{kr93}) and
also the spectroscopy of the nearby nucleus $^{11}$Be suggest a
leading $s_{1/2}^2$ configuration in $^{11}$Li.

In principle, a nuclear-induced breakup gives independent information,
and so it is desirable to calculate the various cross sections and
compare with experiment.  A recent experiment \cite{zi97} was carried
out on a \C12{} target at 280 MeV/n.  At that energy it is justified
to treat the target-projectile interaction in the sudden approximation,
using the NN forward scattering amplitude for the interaction.  
Thus we may neglect
the evolution of the wave function during the interaction time,
provided we take the interaction from nucleon-nucleon scattering. 
The energy domain around 250 MeV has an additional advantage from
a theoretical point of view: the real part of the NN forward scattering
amplitude goes through zero in this vicinity, so only the absorptive part
of the interaction needs to be treated in the theory.  

The nuclear excitation of Borromean nuclei have been considered by a number of
authors \cite{og92,ba93,ba96,th94,ba95,og94,ga96}. In treating the differential
cross sections, it common to make a number of simplifying
assumptions.  We list them here:
\begin{itemize}
\item Ground state wave function. Neutron--neutron correlations were
neglected in ref. \cite{ba93}. We shall apply wave functions that have
the full three-particle correlations.  It turns out that differential
cross sections are quite insensitive to these correlations, except the
two-neutron stripping which does show an effect.  Independent particle
models can only describe pure configurations, so a mixture of {\it s}-
and {\it p}-waves requires a correlated model.  
\item Reaction model.  In this work we use an eikonal model
description of the nuclear reaction, improving on the black disk model
of ref. \cite{ba93}.
\item Neutron--core potential.  It is important to include the
final state neutron-core potential in calculating the energy or
momentum spectra, as demonstrated in ref. \cite{ga96}. 
Ref. [7] made use of simplified
two-body wave functions which were based on a zero range neutron--core
potential.  We shall use a more realistic, finite range potential,
both in the initial and final states.
\end{itemize}
We shall investigate the validity of these as well as other
approximations that are often made.  Our main interest is the
sensitivity of experiments to the properties of the halo nucleus.  In
a previous work \cite{es97} we developed models of the \Li{} ground
state wave function with differing amounts of $s$-wave.  One of our
objectives is to see how well the amount of $s$-wave can be determined by
the observables in a breakup reaction.  The observables we consider
are integrated cross sections for diffraction and one- and two-nucleon
removal, and the differential cross section for the excitation energy
in the $^9$Li + n final state when one neutron has been removed.

\section{Reaction model}
\label{sec:reaction} 

The sudden approximation leads to the eikonal model for nucleus-nucleus
interactions.  In previous studies, we have applied the model to the
nuclear-induced breakup of single-nucleon halo nuclei
\cite{he96}. Here we apply it to the breakup
of a two-neutron halo nucleus.  The effect of the interaction with
the target to multiply the halo wave function by the profile
functions $S(R_i)$ for each particle, where $R_i$ denotes the impact
parameter of particle $i$ with respect to a target nucleus.
The halo nucleus \Li{} has two neutrons and a $^9$Li core, requiring
two types profile functions, $S_n$ and $S_c$, associated with a
neutron and the core, respectively.  There are three integrated cross
sections that leave the core intact, namely the diffractive, the
one-nucleon stripping, and the two-nucleon stripping cross sections.
These can be written
\begin{mathletters}
\begin{eqnarray}
\label{sigma-difa}
\sigma_{dif} &=& 
\int d^2R_{cm} \biggl[\langle \left(1-S_c(R_c)S_n(R_1)S_n(R_2)\right)^2
\rangle - \langle \left(1-S_c(R_c)S_n(R_1)S_n(R_2)\right)\rangle^2\bigg]
\\
\label{sigma-difb}
&=& \int d^2R_{cm} \biggl[\langle S_c^2(R_c)S_n^2(R_1)S_n^2(R_2)
\rangle - \langle S_c(R_c)S_n(R_1)S_n(R_2)\rangle^2\biggr],
\end{eqnarray}
\end{mathletters}
\begin{eqnarray}
\label{1n-st}
\sigma_{1n-st} &=& 2\int d^2R_{cm} \langle S_c^2(R_c)S_n^2(R_2)
(1-S_n^2(R_1)) \rangle, 
\\
\sigma_{2n-st} &=& \int d^2R_{cm} \langle S_c^2(R_c)(1-S_n^2(R_1))
(1-S_n^2(R_2)) \rangle, 
\end{eqnarray}
where $R_{cm}$ is the impact parameter of the halo nucleus with
respect to the target nucleus, and $\langle...\rangle$ denotes a
ground state expectation value. 
Our ground state wave function $\Psi_0$ is expressed in terms of the
relative neutron--core distances, $r_1$ and $r_2$.  An example of the
needed expectation values is the one-nucleon stripping integral
\be
\langle S_c^2 S_n^2(1-S_n^2)\rangle=
\int d^3r_1 d^3 r_2 |\Psi_0(r_1,r_2)|^2 
S_c^2(R_c)
S_n^2(R_c+r_{2\perp})
\Big(1-S_n^2(R_c+r_{1 \perp})\Big) .
\label{sixdim}
\ee
The integrations are here performed for fixed $R_{cm}$ so one should
use $R_c=R_{cm}-(r_{1\perp}+r_{2\perp})/(A_c+2)$, where $A_c$ is the
mass number of the core nucleus.

The six-dimensional integration in eq.~(\ref{sixdim}) is very time
consuming to carry out unless some simplification are made in the wave
function or in the profile functions.  We shall consider two
simplifying approximations.  The first is the {\it no-recoil limit} in
which the impact parameter of the core, $R_c$, is assumed to coincide
with the impact parameter $R_{cm}$ of the halo nucleus.  The core
profile function $S_c$ can then be taken outside the expectation
value.  In addition, the integrations over $r_1$ and $r_2$ become
independent and the six-dimensional integral separates into a product
of two three-dimensional integrals.\footnote{As discussed later on,
the no-recoil limit differs from the exact calculation only in the
case of diffraction.}  Another simplifying approximation is the {\it
transparent limit}, defined here by setting the factor $S^2_n(R_2)$
equal to one inside the expectation value of eq. (\ref{1n-st}), thus
neglecting the absorption of the second neutron.  These two
assumptions yield the cross section
\be
\sigma_{1n-st,trans} = 2 \int d^2 R \ 
S_c^2(R)\langle 1-S_n^2(R+r_{1\perp})\rangle.
\ee
Note that this cross section is identical to the sum of the
one- and two times the two-neutron stripping cross section,
\be
\sigma_{1n-st,trans} = \sigma_{1n-st} + 2 \sigma_{2n-st}.
\ee
We will see later that the two-neutron stripping cross section is
rather small, so the transparent limit is a good approximation for
this cross section.

Of course, much more information about the halo is contained in
differential final state distributions.  The diffractive cross section
has three particles in the final state, but that distribution is
beyond what we can calculate, requiring three-particle continuum wave
functions for many partial waves.  The one-neutron stripping leaves
two particles in the final state, and the differential cross section
for that state is amenable to computation.  The expression for the
momentum distribution associated with the relative motion of the two
surviving particles is
\be
\label{dif-strip-full}
{d \sigma_{st} \over d^3 k}
= 2 \int d^2R_1 \left[1-S^2_n(R_1)\right]
\int d^3r_{2c} \Bigl| M(R_1,r_{2c},k) \Bigr|^2,
\ee
where $r_{2c}$ is the center-of-mass coordinate of the remaining
neutron--core system with respect to the stripped neutron; the
associated impact parameter with respect to the target nucleus is
denoted by $R_{2c}$, $R_{2c}=R_1+r_{2c\perp}$.  The amplitude $M$ is
given by
\be 
M = \int  d^3r_2 \ \psi_c^*(k,r_2)
S_c\big(R_c\bigr)
S_n(R_2)\Psi_0(r_1,r_2) .
\label{mstrip}
\ee
Here $\psi_c(k,r_2)$ is the continuum wave function of the surviving
neutron--core system, normalized to a plane wave at infinity.  The
coordinates $R_c$, $R_2$ and $r_1$ are expressed in terms of the
integration variables as $R_c=R_{2c}-r_{2\perp}/(A_c+1)$,
$R_2=R_{2c}+r_{2\perp}A_c/(A_c+1)$, and $r_1=-r_{2c}+r_2/(A_c+1)$.

The numerical calculation of eq.~(\ref{mstrip}) is rather difficult
because of the form of the ground state wave function that we apply
(see the next section). A major simplification is achieved by adopting
the approximation $r_1$ = $-r_{2c}$ in the ground state wave function.
The amplitude is then given by
\be
\label{appm'}
M' = \int  d^3r_2 \ \psi_c^*(k,r_2) S_c(R_c)S_n(R_2)
\Psi_0(-r_{2c},r_2) . 
\ee
An even simpler approximation is to also ignore the recoil correction
in the argument of the core profile function, i.~e. setting
$R_c=R_{2c}$, and to use the transparent limit for the second neutron,
i.~e.  setting $S_n(R_2)=1$. We shall refer to these approximations as
the {\it no-recoil, transparent limit}, where the amplitude reduces to
\be
\label{appm''}
M'' = S_c(R_{2c}) \int  d^3r_2 \ \psi_c^*(k,r_2) \Psi_0(-r_{2c},r_2) . 
\ee
This approximation is used in ref. \cite{ba96} and \cite{ga96}. We will
discuss the validity of the various approximations in 
Sec.~\ref{sec:stdist} below.

\section{The three-body wave function}
In ref. \cite{es97} we constructed several three-body models of \Li.
The models are based on Hamiltonians that all reproduce the empirical
neutron--neutron scattering length and all have a binding energy of
the three-body system close to the empirical value of 295 $\pm$ 35 keV
\cite{yo93}.  The single-particle potentials and the
density-dependence of the neutron--neutron interaction are varied to
produce different probabilities of $s$- and $p$-waves in the different
models.  We will specifically examine the observables for models
having 4.5\%, 23\%, and 50\% $s$-wave.  The first model is similar to
the one used in ref. \cite{be91}.  The second and third models are
constructed with a deeper neutron--core potential for even parity,
single-particle states to increase the $s$-wave component in the
ground wave function. Details of the procedure and the first two models 
are given in ref. \cite{es97}.  The wave functions are calculated in the 
form of single-particle states $u(r)$ and amplitudes $\alpha$ as
\be
\psi_0(\vec r_1,\vec r_2) = 
\sum_{lj} \sum_{n,n'}\alpha_{ljnn'} u_{ljn}(r_1) u_{ljn'}(r_2)
\Big[(l s)^j (l s)^j\Big]^0
\ee
where $\vec r_1$ and $\vec r_2$ are neutron--core separation vectors.
The indices $(ls)^j$ label the single-particle, spin-angle wave
functions which are coupled to zero total angular momentum as
indicated by the superscript on the bracket.  The indices $n,n'$ label
the radial quantum numbers of the single-particle basis states.  These
states are discretized by putting the system into a spherical box of
large radius (typically 40 fm).

\begin{figure}
\vspace*{\fill}
\begin{center}
\ForceHeight{10cm}
\BoxedEPSF{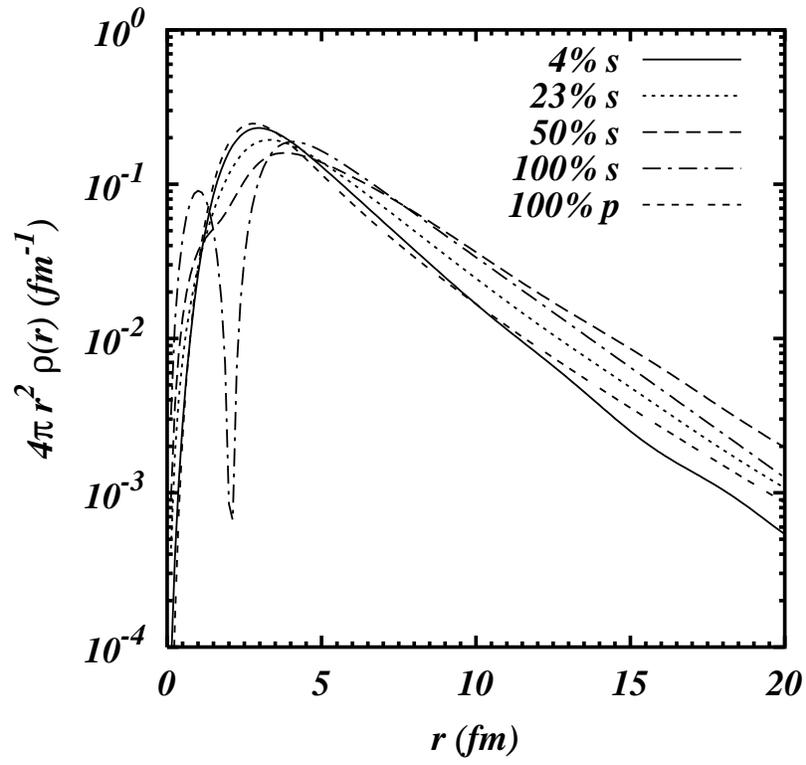}
\end{center}
\vspace*{\fill}
\caption{
Single-particle density of halo neutrons in various models of
\Li.
}
\label{fig:1}
\end{figure}

In Table~\ref{tab:models} we give some characteristics of the
Hamiltonians and the resulting ground state wave-functions.  A useful
quantity is the $s$-wave scattering length.  Within the constraints of
our three-body model, we can only produce wave functions that are
predominantly $s$-wave by using neutron--core potentials that produce
extremely large $s$-wave scattering lengths.  The parameters of the
potential for the odd-parity states are fixed by the position of the
$p_{1/2}$ resonance, which we assume to be at $E_r=540$~keV as
suggested by measurements \cite{yo94}.  Besides the $s$-wave
probability, these wave functions show significant differences. The
single-particle densities of the three models are shown in Fig.~\ref{fig:1}. 
It
may be seen that the halo is more extended the larger the $s$-wave
probability.  This is also apparent from the mean square neutron radii
computed in Table~\ref{tab:models}.

The integrated dipole strength for Coulomb breakup is proportional
to the mean square radius of the two-neutron center of mass, given
in the last column of Table I.  The value obtained with the p89
wave function is consistent with the experimental Coulomb breakup,
but the s50 value is much too high.  Thus, we cannot regard that
wave function as realistic.  

Another important property of the wave function is the correlation
between the two neutrons.  The integrated dipole strength is
proportional to the dineutron--core mean square radius, which in turn
depends on the matrix element of $\vec r_1 \cdot \vec r_2$, as shown
in \cite{be91}.  In that work it was found that the correlation
increased the dipole strength by 43 \%.  This enhancement does not
depend very much on the model; with the present wave functions it is
in the range of 30-40\%, which may be extracted from the mean
square radii computed in Table~\ref{tab:models}.

We also constructed $s$- and $p$-wave independent-particle models for
comparison purposes.  In these models, the single-particle potential
is adjusted to match the exponential fall-off of the single-particle
density that is obtained with the three-body model s23 mentioned in
Table~\ref{tab:models}.  We see from Fig.~\ref{fig:1} that the pure $s$-wave
model gives a much more diffuse halo than the $p$-wave model or the
correlated models.

\section{Profile functions}
We now specify the profile functions $S_n$ and $S_c$ that we use for
our cross section calculations.  The neutron profile function $S_n$ in
the eikonal approximation is expressed in terms of the density of the
target $\rho_t$ and the nucleon-nucleon cross section $\sigma_{nn}$ as
\be
S(b) = \exp \Bigl[ -{\sigma_{nn}\over 2}\int dz \rho_t(\sqrt{b^2+z^2})
\Bigr]].
\ee
We model the density of the \C12{} target with the harmonic oscillator
fit to the charge density of ref. \cite{de74},
\be
\label{rhoho}
\rho(r) = \rho_0 \Bigl[ 1 + \alpha (r/a)^2\Bigr] e^{-(r/a)^2},
\ee
with $a = 1.687$ fm and $\alpha = 1.067$.  The nucleon-nucleon cross
section is taken from ref. \cite{ch90}; it is 29.2 mb at 280 MeV beam
energy.

\begin{figure}
\vspace*{\fill}
\begin{center}
\ForceHeight{10cm}
\BoxedEPSF{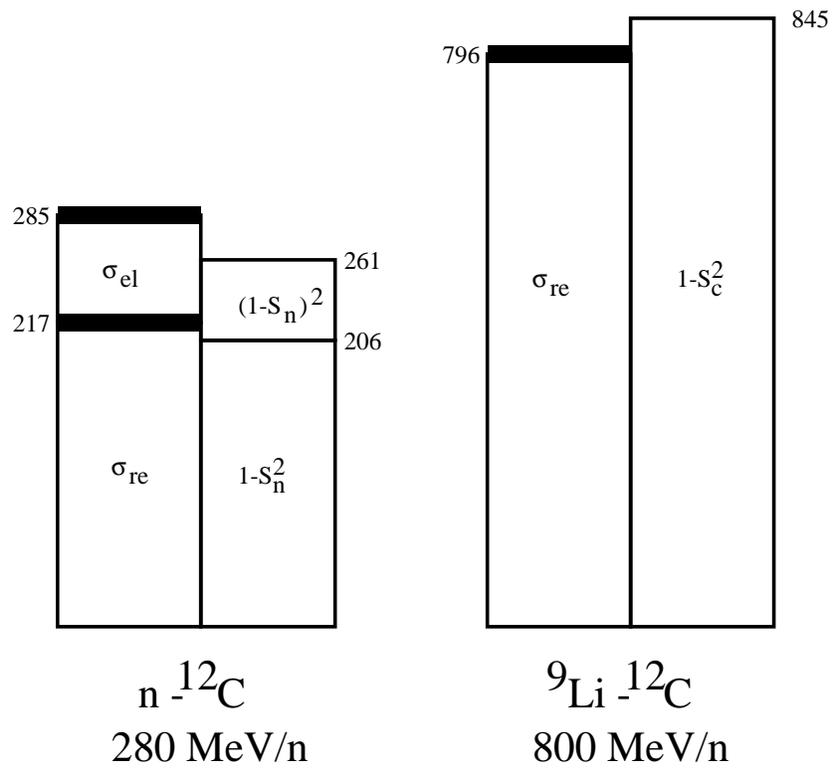}
\end{center}
\vspace*{\fill}
\caption{
Cross sections for nucleon-\C12{} scattering and for $^9$Li-\C12{} scattering. 
Comparison of the results of our profile function with experimental results.
}
\label{fig:2}
\end{figure}

The reliability of the model can be checked against the nucleon-carbon
cross sections.  The predicted reaction and elastic cross sections in
the eikonal model are
\begin{eqnarray}
\sigma_{re} &=& \int d^2b \Bigl[1-S_n^2(b)\Bigr],
\\
\sigma_{el} &=& \int d^2b \Bigl[1-S_n(b)\Bigr]^2.
\end{eqnarray}
These are compared with experiment in Fig.~\ref{fig:2}. The nucleon-carbon
reaction cross section is taken from the proton cross section data
tabulated in ref. \cite{ba88}, quoting ref. \cite{re72}.  The total
cross section for nucleon-carbon scattering is taken from the neutron
measurements of ref. \cite{fr88}.  The experimental elastic cross
section is deduced from the difference between total and reaction
cross section.  The agreement between our parameterization of $S_n$
and experiment is close enough that we will not attempt to adjust the
profile function to make a better fit.  In the next section we will
discuss how the cross sections in halo nuclei depend on the
nucleon-target cross sections.

The core-target profile function requires the convolution of both
densities,
$$
S_c(b) = \exp \Bigg[ -{\sigma_{nn}\over 2} \int dx dy
\int dz \rho_c\Big(\sqrt{(x-b)^2+y^2+z^2}\Big)
\int dz' \rho_t\Big(\sqrt{x^2+y^2+z'^2}\Big)\Bigg].
$$
For the density of $^9$Li, we note that it has the same number of
neutrons as \C12{} and we will accordingly take the same parameters
for the neutrons. The proton density does not have as many particles
in the $p$ shell, and we apply the pure harmonic oscillator model to
determine $\alpha$ ($=1/3$), and keep $a$ the same as in \C12. The
resulting $^9$Li density is parameterized as in eq.~(\ref{rhoho}) with
$a=1.687$ and $\alpha=0.726$.  This model gives an rms charge radius
of 2.28 fm, slightly smaller than the empirical charge radius of
$^7$Li which is 2.39 fm.  However, the predicted cross section at 800
MeV/n is 840 mb (assuming $\sigma_{nn}= 40$ mb), just 5\% larger than
the measured cross section of $796\pm6$ mb from ref. \cite{ta85}.  The
cross section at 280 MeV/n has been measured for the mirror nucleus
$^9$C by Blank et al. \cite{bl97}.  They find a cross section of $812
\pm 34$ mb to be compared with 796 mb obtained by our model.

For the $^4$He core of \He6{} we use a 3-parameter-Fermi density
function \cite{de74},
$$\rho_c(r)=(1+w r^2/c^2)/\left[1+\exp((r-c)/z)\right],$$ with
$w=0.517$, $c=0.964$~fm, and $z=0.322$~fm.  At 800~MeV/n we find a
total cross section of 546~mb (486 at 280 MeV/n) again comparable with
the experimental result of $503 \pm 5$~mb
\cite{ta85a}.

\section{Integrated cross sections}
In this section we examine the integrated cross sections and compare
with experimental data.  It is important to understand the dependence
of the cross sections on the interaction and on the properties of the
halo, and to this end we first discuss some estimates and bounds on
the cross sections.

\subsection{Cross section bounds}
\begin{figure}
\vspace*{\fill}
\begin{center}
\ForceHeight{8cm}
\BoxedEPSF{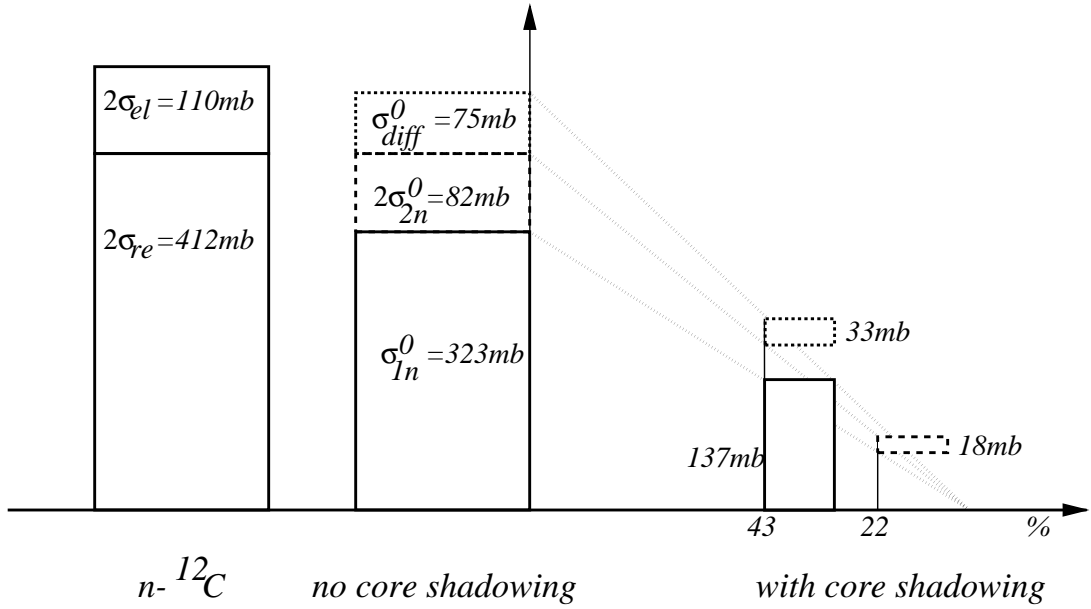}
\end{center}
\vspace*{\fill}
\caption{
Breakup reaction cross sections for \Li-\C12{} scattering. The 
left-hand histogram shows twice the $n$-\C12{} cross sections as required by
equality eq.~(\protect\ref{strip-noshadow}) and the bound for the diffractive cross
section, eq.~(\protect\ref{difelb}). We compare them with the total cross sections
without core shadowing (``$\sigma^0$'') in the next histogram. The shadowing
reduces the cross section by a factor of $\approx0.4$ for
single-neutron stripping and diffraction, and by a factor of
$\approx0.2$ for two-nucleon stripping, as show by the diagram on the
right side.
}
\label{fig:3}
\end{figure}

Let us first consider the (unrealistic) limit where the core-target
interaction is ignored\footnote{This is commonly referred to as the transparent limit but
we have reserved that concept for the transparency of the second
neutron in a 1n-stripping reaction, c.~f. Sec.~\ref{sec:reaction}.}.  
For a
single-nucleon halo the stripping cross section would then be
identical to the reaction cross section $\sigma_{re}$ between the
nucleon and the target.  In the case of a two-nucleon halo, the
stripping cross section is just doubled.  Split into 1n- and
2n-stripping components, the relation is
\be
\label{strip-noshadow}
\sigma_{1n-st}^0 +2 \sigma_{2n-st}^0 = 2 \sigma_{re}.
\ee
The symbol $\sigma^0$ is a reminder that the core shadowing is
neglected, i.e. the factor $S_c$ is set equal to one.
Eq. (\ref{strip-noshadow}) is illustrated in Fig.~\ref{fig:3}, showing the
comparison of the left- and right-hand sides of the equation for the
case of \Li{} breakup on a \C12{} target.  We used the s23 model to
evaluate the unshadowed cross section.  The relative amounts of 1n-
and 2n-stripping depends of course on the wave function and details of
the interaction; in the case considered here, the 1n-stripping cross
section is an order of magnitude larger than the 2n stripping.

The diffraction cross section is much more difficult to bound or
estimate without full calculation of the integrals.  In the case of a
one-nucleon halo, a bound can be obtained by dropping the second term
in the equation analogous to eq. (\ref{sigma-difa}).  The first term
is just the elastic nucleon-target cross section, so the bound is
$\sigma^0_{dif}\le \sigma_{el}$.  For $^{11}$Be at 800~MeV/n and using
the model of \cite{he96} we get $\sigma^0_{dif}=68$~mb, smaller than
the bound $\sigma_{el}=85$~mb.  This shows that the neglected term is
not necessarily small.  For the two-nucleon diffraction formula
eq. (\ref{sigma-difa}) the first term does not reduce to the elastic
scattering, so no strict bound can be obtained this way.

Instead, we shall analyze diffractive cross section qualitatively
using a grey-disk model for the nucleon-target interaction.  Thus, we
take the nucleon profile function to be of the form $S_n(b) = 1-t
\theta(b_0-b)$, $0\le t\le 1$.  We also need to assume that the two
nucleons are uncorrelated in the halo wave function.  The diffraction
cross section may then be expressed as
\be
\label{dif-noshadow}
\sigma_{dif}^0 = 2 \sigma_{el}\Bigl[1-(1+2t-t^2/2)a\langle\rho_t\rangle
+ 2ta^2\langle\rho_t^2\rangle
-{t^2\over 2}a^3\langle\rho_t^3\rangle\Bigr],  
\ee
where $\langle\rho_t^n\rangle$ is the $n$-th moment of the transverse
nucleon density in the halo\footnote{More precisely, it is the $n$-th
moment of an averaged transverse density, the averaging being over the
shape of the nucleon profile function.}  and $a=\pi b_0^2$. The
leading term is twice the nucleon-target elastic cross section and the
corrections are controlled by the parameter
$a\langle\rho_t\rangle$. The coefficient of the first correction is
negative, so the first term gives a bound that is valid for large
halos and small targets,
\be
\label{difelb}
\sigma_{dif}^0 \le 2 \sigma_{el}.
\ee
In the case of a \C12{} target, the experimental elastic and reaction
cross sections may be fit with $a=30 $ fm$^2$ and $t=0.5$.  The
transverse halo density for the \Li{} wave functions has the order of
magnitude $\langle\rho_t\rangle\approx 1/100$ fm$^{-2}$ and the second
term makes about a factor of two correction; the higher terms are less
important.  The actual numbers for our model of the \Li{}--\C12
reaction are shown in Fig.~\ref{fig:3}.  We find $\sigma^0_{dif}=75$ mb, reduced
from $2\sigma_{el}$ by about a factor $2/3$, as expected from the
above analysis.

\subsection{Core shadowing}
In this section we examine the effect of the core shadowing on the
cross sections, and use again the model s23 for the ground state wave
function. Fig.~\ref{fig:3} shows on the right the shadowing effect 
of the carbon target in the \Li{} breakup reaction. The 1n-stripping
cross section is reduced to 43\% of the unshadowed value
$\sigma^0_{1n-st}$.  The shadowing factor for the diffractive cross
section is very similar (44\%) to that for the 1n-stripping.

\begin{figure}
\vspace*{\fill}
\begin{center}
\ForceHeight{10cm}
\BoxedEPSF{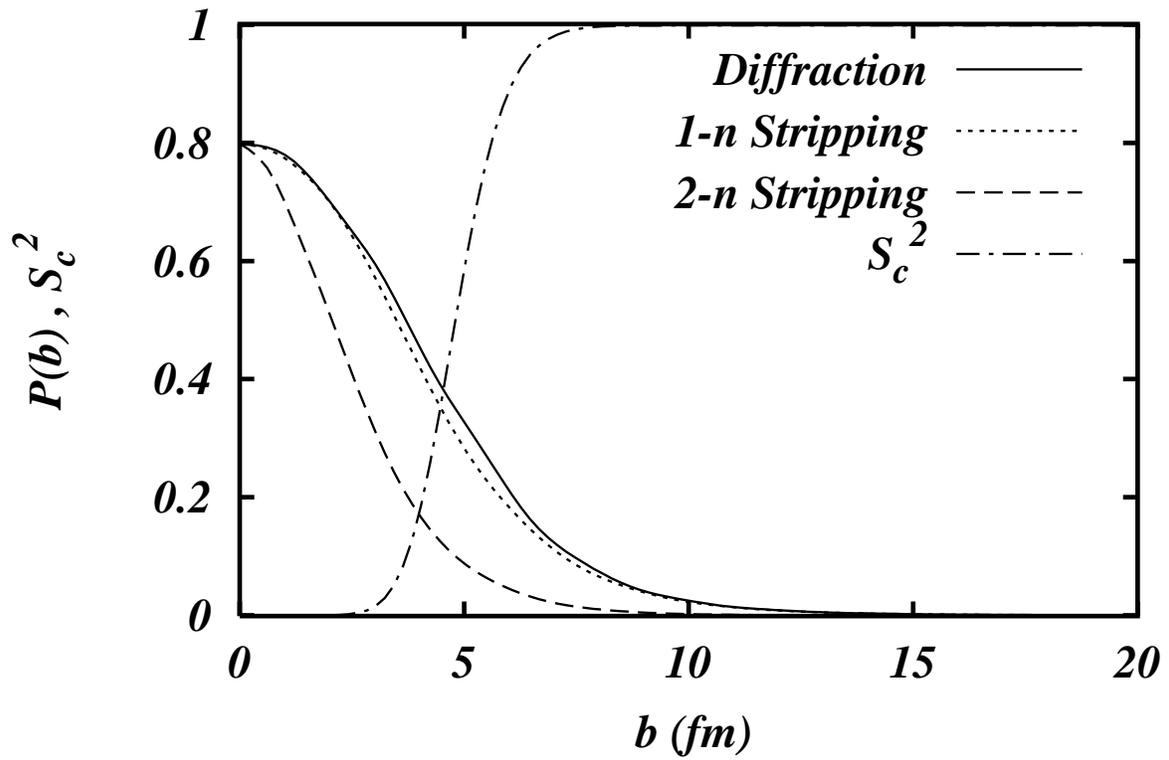}
\end{center}
\vspace*{\fill}
\caption{
Impact parameter dependence of the 1n- and 2n-stripping and
the diffractive cross sections.  Also shown is the square of the profile
function of the core-target interaction, $S_c^2$.
}
\label{fig:4}
\end{figure}

The shadowing factor for two-nucleon stripping is much stronger than
for the other processes; it reduces the cross section to 20\% of the 
unshadowed value.  The difference may be
understood qualitatively as follows.  The one-nucleon stripping and
the diffractive excitation require avoiding an absorptive interaction
with a least one of the halo nucleons, favoring moderately large
impact parameters.  On the other hand, the two-nucleon absorption has
no such restriction and would be concentrated entirely at small impact
parameters but for the presence of $S_c$.  The different dependences
on impact parameter is shown in Fig.~\ref{fig:4}.  Here we see that the
2n-stripping probability is more concentrated at small impact
parameter than the 1n-stripping and the diffractive probabilities,
which are very similar to each other.

\begin{figure}
\vspace*{\fill}
\begin{center}
\ForceHeight{10cm}
\BoxedEPSF{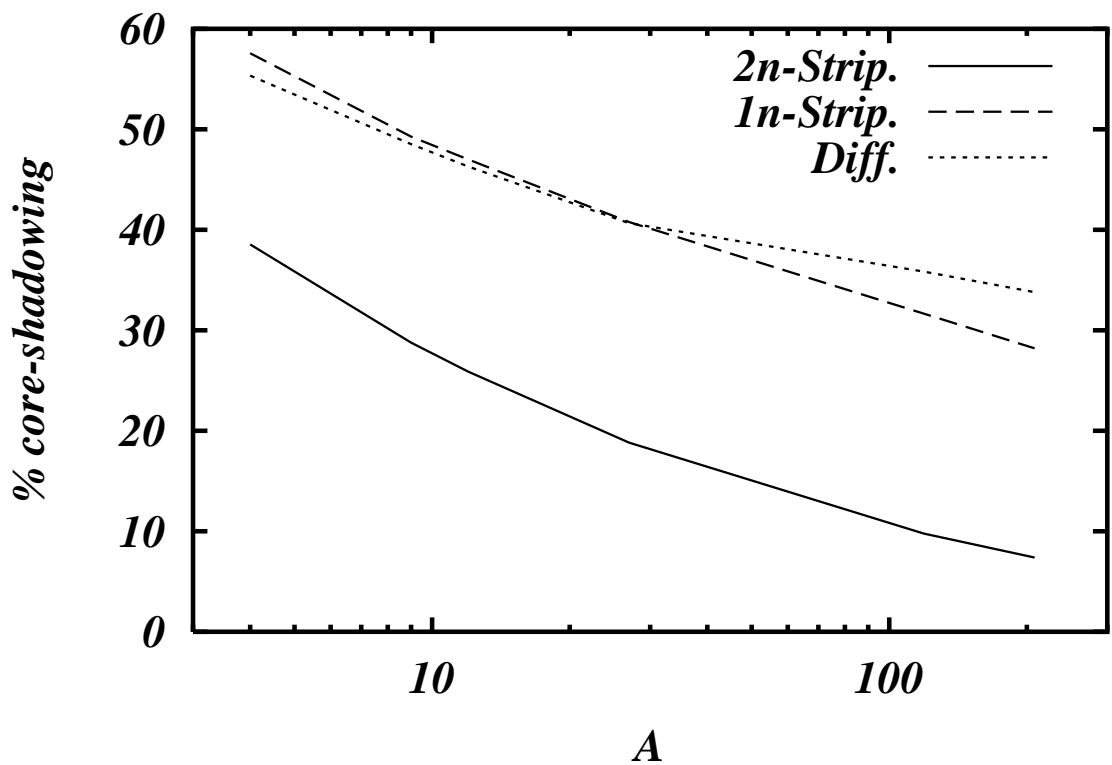}
\end{center}
\vspace*{\fill}
\caption{
Shadowing factor as a function of target mass number.
Shadowing factors are shown for 2n-stripping, 1n-stripping, and
diffractive cross sections at an energy of 280 MeV/n.
}
\label{fig:5}
\end{figure}

The shadowing factor varies, of course, with target size.  This
dependence is illustrated in Fig.~\ref{fig:5}, where the target densities were
taken from \cite{de74,de87}.  We see that the shadowing changes by a
factor of 2 both for diffraction and 1n-stripping, going from a $^4$He
target to a heavy target, and by a factor of 4 for the 2n-stripping.

\subsection{Wave function sensitivity}
We next consider the sensitivity of the cross sections to properties
of the halo wave functions.  The various cross sections for different
models are given in Table~\ref{tab:xs}.  For the single-nucleon
stripping cross section, the shadowing factor varies, depending on how
extended the single-particle density is.  From Table~\ref{tab:models},
we see that the mean square radius of the halo increases as the
$s$-wave probability increases.  Thus we expect less shadowing and a
larger cross section for the models with a larger $s$-wave.  This is
indeed born out by the numbers in Table~\ref{tab:xs}.  For the
two-nucleon stripping, the correlation between the neutrons should be
important as well, as they must both interact with the target.
Indeed, we see from Table~\ref{tab:xs} that the two-nucleon cross
sections doubles going from an uncorrelated $p$-wave model to the
model with correlations and $p$-wave dominance.  This may be compared
with the effect of the correlations on the dipole transition strength
which, as was mentioned in the last section, gives only a 30-40\%
enhancement.

\subsection{\He6{} cross sections}
Here we report corresponding cross sections
for the breakup of \He6{} on a \C12~target, using the \He6{}
wave function from ref. \cite{es97}, line 5 of Table II.
The cross sections for two different beam energies are shown in
Table~\ref{tab:he6}. \He6{} is more tightly bound than \Li, so the 
halo density
does not extend out as far.  Another difference is that \He6{} has a
dominant $p_{3/2}$ shell configuration which allows a stronger spatial
correlation; pure $s_{1/2}$ or pure $p_{1/2}$ configurations, on the
other hand, have uncorrelated densities.  The larger correlation
implies that the 2n-stripping will be relatively stronger.  This is
indeed seen to be the case in Table~\ref{tab:he6}; the 2n-stripping
cross sections is about a factor of two larger for \He6{} than for
\Li.  Otherwise, the cross sections are about the same as for \Li.
The shadowing factors are similar, due to balancing features of a
smaller core and a less extended halo.

\section{Shape of the stripping spectrum}
\label{sec:shape}
In this section we discuss analytic forms for the shape of the
spectrum in the neutron--core system produced by the 1n-stripping
reaction. The standard parameterization of a peaked distribution by
the Breit-Wigner function is not justified at energies close to zero,
or for the overlaps with extended wave functions.  We shall propose
parameterizations that take into account the threshold behavior and
the halo character of the initial state.  Throughout this section we
make use of the no-recoil, transparent limit defined in
eq. (\ref{appm''}).

To treat the one-neutron removal from a Borromean nucleus, we simply
take the overlap of the initial ground state wave-function with the
continuum final state of the neutron--core system
(cf. eq. (\ref{appm''})), fixing the position $r_1=-r_{2c}$ of the
stripped neutron.  The stripping model assumes that the process is
incoherent in $r_1$.  Thus we consider matrix elements of the form
\be
\label{appmk}
M''(r_1,k) = \int d^3 r_2 \ \psi_k^*(r_2) \psi_0(r_1,r_2)
\ee
and a probability distribution of the form 
$$
|M''(r_1,k)|^2 d n_k.
$$
Here $d n_k\sim k^2 dk$ for a differential momentum distribution and
$d n_k \sim k d E$ for a differential distribution in excitation
energy of the neutron--core system.

Different partial waves of the continuum wave function are incoherent
if we integrate over the direction of the decay distribution.  At low
energies the $s$- and $p$- waves will be most important, and we now
discuss their functional behavior.

\subsection{$s$-Wave distribution}
The $s$-wave distribution can be described analytically in the limit
where the wave functions are dominated by their asymptotic
behavior\footnote{The formula derived gives an excellent description
of the magnetic photo-disintegration of the deuteron at low
energy\cite{de64}.}.  The continuum $s$-wave is then given by
$$
\psi_0 \sim {\sin(k r + \delta) \over k r}.
$$
Here $\delta$ is the $s$-wave phase shift; the scattering length is
the linear coefficient in the expansion
$$
\delta = -a k + O(k^3).
$$
The two-particle initial state wave function has no exact analytic
limit, but the general exponential fall-off at large distances
suggests the approximation
$$
\psi(r_2,r) \approx f(r_2) {e^{-\alpha r}\over r}.
$$
Then the integral in eq. (\ref{appmk}) can be carried out to give
$$
M''\sim {\cos(k a) -\alpha/k \sin ( k a)\over k^2 + \alpha^2},
$$
and therefore the cross section for the $s$-wave is given by
\be
\label{s-analytic}
\frac{d\sigma}{dE} \sim  k \left[ \frac{1}{\alpha^2+k^2}\right]^2
\left[ \cos(ka) - \frac{\alpha}{k} \sin(ka) \right]^2.
\ee
This depends on the initial state potential through the fall-off
parameter $\alpha$ and on the final state potential through the
scattering length $a$.  If the two potentials are the same, then
the orthogonality of initial and final states requires the matrix
element to vanish.  This comes about in eq. (20) to leading order
in $k$ by the well-known relation between binding energy and
scattering length \cite{el59}.

\begin{figure}
\vspace*{\fill}
\begin{center}
\ForceHeight{10cm}
\BoxedEPSF{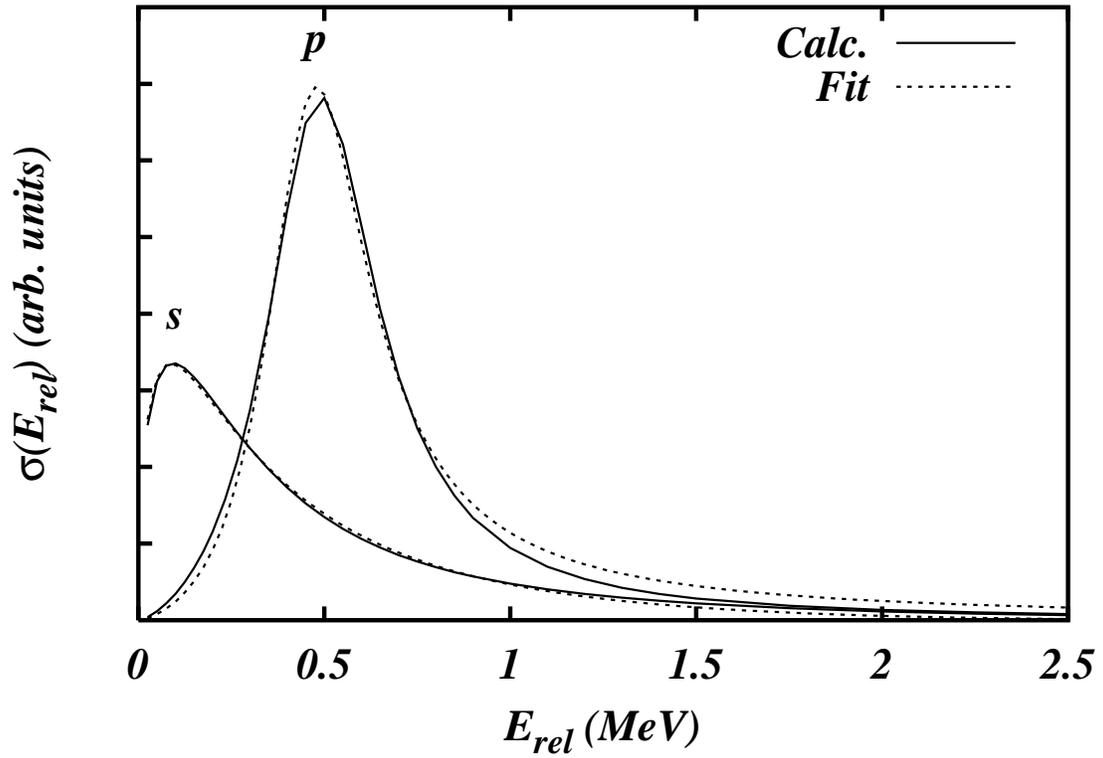}
\end{center}
\vspace*{\fill}
\caption{
The \Li{} 1n-stripping energy distributions in the no-recoil,
transparent limit (solid curves) are compared to fits obtained from
eq. (\protect\ref{s-analytic}) and eqs. (\protect\ref{p-analytica},\protect\ref{p-analyticb})
(dashed curves). Shown are the $s$- and $p$-wave components of the s23
wave function.
}
\label{fig:6}
\end{figure}

The $s$-wave energy distribution for \Li{} is shown in Fig.~\ref{fig:6}.  Here we have fitted
both parameters $\alpha$ and $a$ to give the best agreement with the
calculated curve, which was obtained in the no-recoil, transparent
limit (cf. eq. 10) using the model s23. In principal $\alpha$ is given
by the binding energy as $\alpha\approx \sqrt{2 m E_B}$.  Our fit has
$\alpha=24.5$~MeV/c.  The corresponding binding energy is 0.32~MeV,
almost equal to the binding energy of the independent particle
model. Also our fitted value $a=-4.7$~fm is very close to the actual
scattering length of $-5.6$~fm.

The distribution in Fig.~\ref{fig:6} peaks at very low energies; the peak
position is close to the momentum $k=\alpha/2$ for a fairly wide range
of scattering lengths $a$ between $-1/\alpha<a<1/\alpha$. This corresponds
to an energy peak $E_{peak}$ at
$$  E_{peak} = E_B/4. $$
With our theoretical fit, $E_{peak}\approx 0.08$ MeV.
It should also be mentioned that for models with very
large scattering lengths, such as the s50 model, the scattering
length sets the momentum scale and the predicted peak is lower in
energy.

\begin{figure}
\vspace*{\fill}
\begin{center}
\ForceHeight{10cm}
\BoxedEPSF{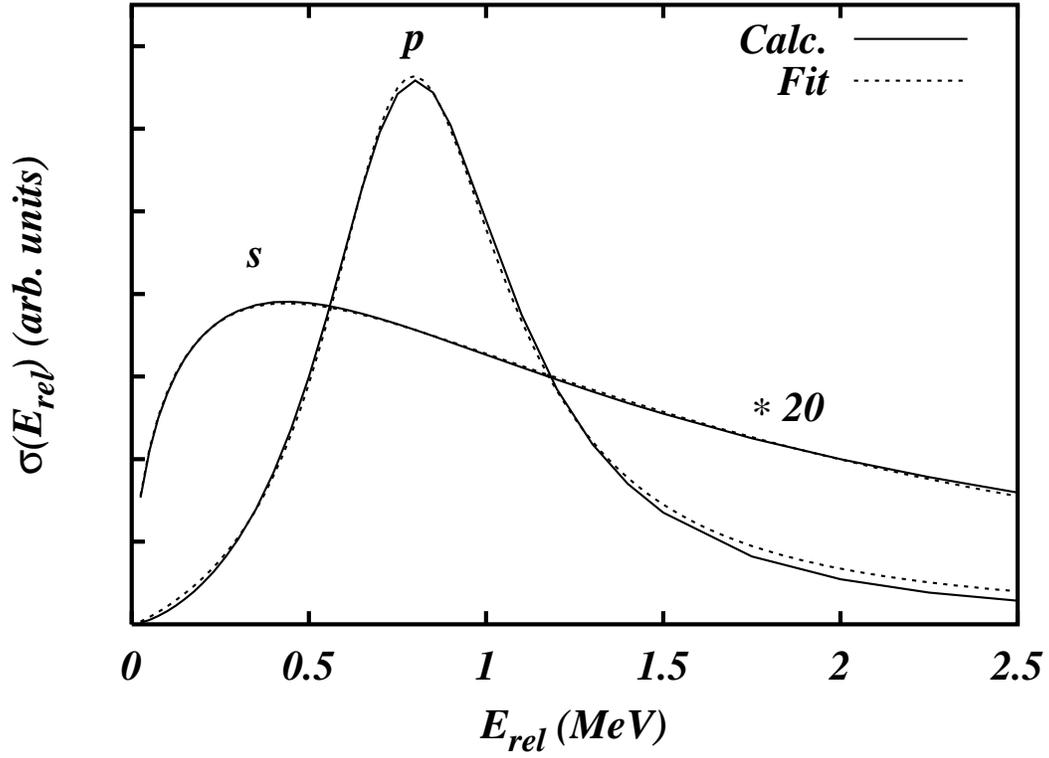}
\end{center}
\vspace*{\fill}
\caption{
The \He6{} 1n-stripping energy distributions calculated in the
no-recoil, transparent limit (solid curves) are compared to fits
obtained from eq.~(\protect\ref{s-analytic}) and
eqs.~(\protect\ref{p-analytica},\protect\ref{p-analyticb}) (dashed curves). The
$s$-wave components has been scaled by a factor of 20.
}
\label{fig:7}
\end{figure}

The corresponding $s$-wave distribution for \He6{} is shown in Fig.~\ref{fig:7}.
Here the best scattering length parameters is $a=1.6$ fm, to be
compared with the actual scattering length of $a=2.4$ fm associated
with the $^5$He potential.  The best fit value of $\alpha$ is
$\alpha=55$ MeV/c; this may be compared with the binding energy
estimate $\sqrt{2 m E_B} = 43$ MeV/c.

\subsection{$p$-Wave distribution}
For the $p$-wave, measurements of the $^{11}$B($^7$Li,$^8$B)$^{10}$Li
reaction have suggested the existence of a resonance at about 540 keV
\cite{yo94}. In our recent study of the \Li{} wave function we used 
this data to fix the $p$-wave potential for the neutron--core system.
In this section we wish to establish a simple function to represent
the distributions that we calculate.  After trying different
functional forms, we found that one could get acceptable fits with the
Breit-Wigner resonance form but using an energy-dependent width.  The
threshold behavior of a $p$-wave resonance requires a width depending
on energy as $\Gamma \sim E^{3/2}$.  However, the width cannot
continue to grow as the $3/2$ power at energies above the resonance.
We shall account for this by using the form of the $p$-wave width
obtained in potential scattering\cite {bm69}
\be
\label{p-analytica}
\Gamma =\Bigl({E\over E_R}\Bigr)^{3/2}  {\Gamma_0 \over 1+c E/E_R}.
\ee
The Breit-Wigner function for the decay of a resonance is
then given by 
\be
\label{p-analyticb}
{d \sigma \over dE} = A { \Gamma \over (E-E_R)^2 + \Gamma^2/4}.
\ee
There are four parameters here, namely the resonance energy $E_R$, the
width on resonance $\Gamma_0$, the overall strength $A$, and a cutoff
parameter $c$.  One might think that $\Gamma_0$ and $c$ could be determined
by the radius of the potential forming the resonance, but the fact
that we are considering a transition from a halo state implies that
the length scales are larger than the nuclear radius.  We shall treat
them as adjustable parameters.  Fig.~\ref{fig:6} shows a fit with parameter
values $\Gamma_0=1$ MeV, $E_R = 0.51$ MeV, and $c=1.7$. When we make an 
unconstrained fit, the parameters $\Gamma_0$ and $c$
acquire large values, showing that the function $\Gamma$ is close to
the $E^{1/2}$ dependence, except at extremely low energies.  The deviation
from the 3/2 power law is quite understandable; although the
single-particle wave function is in a $p$-wave, the fact that the
initial state is a halo means that the extension of the wave
function rather than the radius of the barrier sets the scale for the
threshold region. 

In Fig.~\ref{fig:7} we show similar a comparison for \He6{} stripping.  In this
case, the peak of the $p$-wave distribution is located at 0.83 MeV
which corresponds quite well to the resonance energy of the $p_{3/2}$
scattering state (0.89~MeV).  Nevertheless, the best fit again favors
large values of $c$, indicating that the 3/2 power law for the width
is only valid very close to the threshold.

\section{Stripping Distributions}
\label{sec:stdist}
In this section we treat the full stripping cross section as defined
by eqs.~(\ref{dif-strip-full}) and (\ref{appm'}).  There are a number
of questions one can ask: i) How reliable is the transparent limit to
deduce the shape of the differential cross section? Do we get the
right shape for the $s$ and $p$-wave and also the right ratio between
them?  ii) Can a decomposition of the differential cross section into
$s$- and $p$-wave components be used to infer the $s/p$ ratio in the ground
state wave function?

\begin{figure}
\vspace*{\fill}
\begin{center}
\ForceHeight{10cm}
\BoxedEPSF{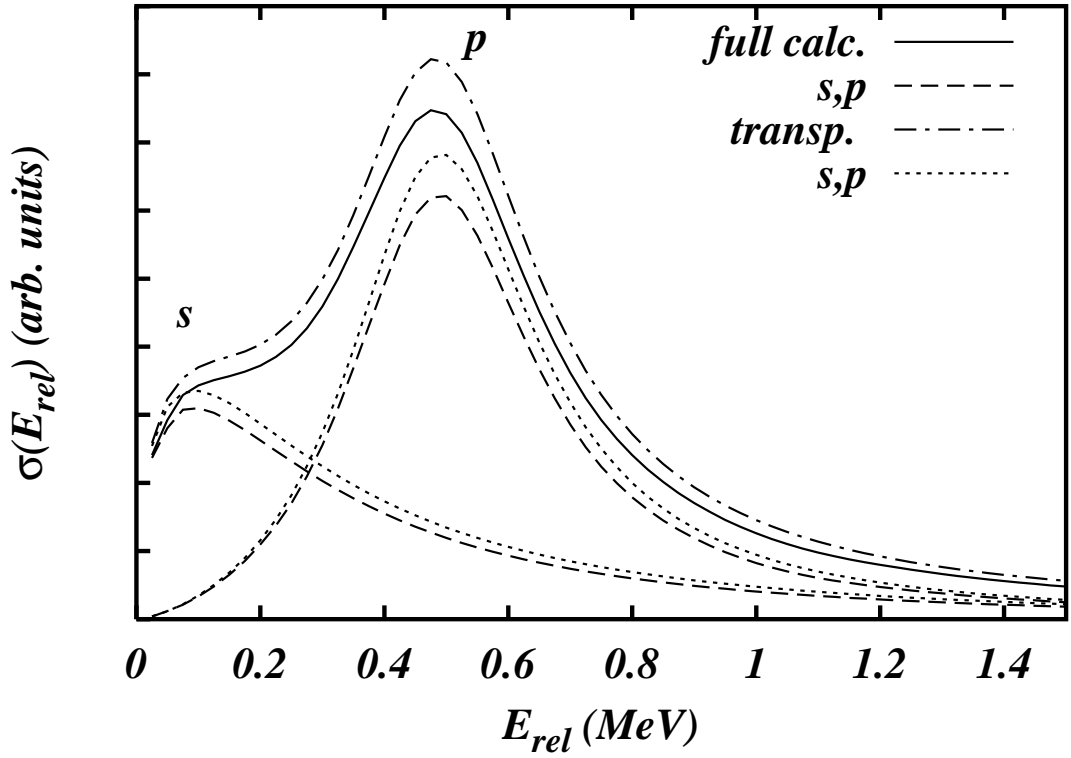}
\end{center}
\vspace*{\fill}
\caption{
Comparison of the full calculation (including shadowing of the
spectator neutron) with the transparent limit. Shown are the total
1n-stripping energy distribution as well as the $s$- and $p$-wave
components of the s23 model.
}
\label{fig:8}
\end{figure}

The energy distribution for the full calculation is compared in Fig.~\ref{fig:8}
to the no-recoil, transparent limit for \Li{} stripping, using the s23
ground state wave function.  We see that the effect of the neutron
shadowing is to reduce the cross section without affecting the energy
distribution.  Thus we can use the transparent limit with confidence
in describing these distributions.

\begin{figure}
\vspace*{\fill}
\begin{center}
\ForceHeight{10cm}
\BoxedEPSF{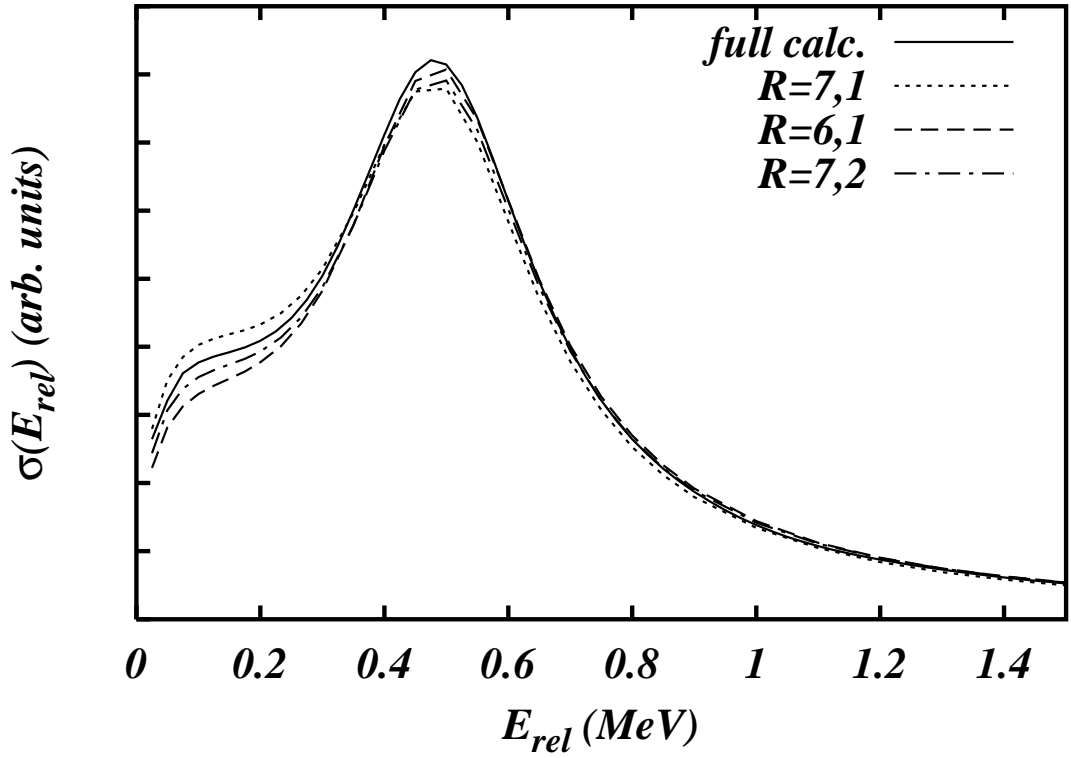}
\end{center}
\vspace*{\fill}
\caption{
Comparison of the 1n-stripping energy distribution of \Li{}
with calculations with fixed impact parameters of the nucleon-core
system (6 and 7fm) and the stripped neutron (1 and 2 fm). The
different spectra have been normalized to the same total value.
}
\label{fig:9}
\end{figure}

The 1n-stripping calculation involves integrations of the probability
over the impact parameters of the absorbed neutron and the center of
mass of the second neutron with respect to the core.  These integrals
are incoherent, and can be performed on a coarser mesh than the wave
function integral over the internal coordinate.  It is of interest to
see how much the shape depends on the impact parameters, to make this
integration as coarse as possible.  Fig.~\ref{fig:9} shows the distribution with
the s23 wave function, comparing the full calculation with the
distributions at typical impact parameters. Here we fix the value of
the transverse radii $R_{2c}$ and $R_1$ and integrate over the angle
between them, and also over the longitudinal component of $r_{2c}$.
The curves have been normalized to the same area.  We see that the
ratio of cross sections at low energy to that at the $p$-wave peak
varies by the order of 20\%.  Thus the impact parameter integration
mesh cannot be coarser than 1 fm or so.

\begin{figure}
\vspace*{\fill}
\begin{center}
\ForceHeight{10cm}
\BoxedEPSF{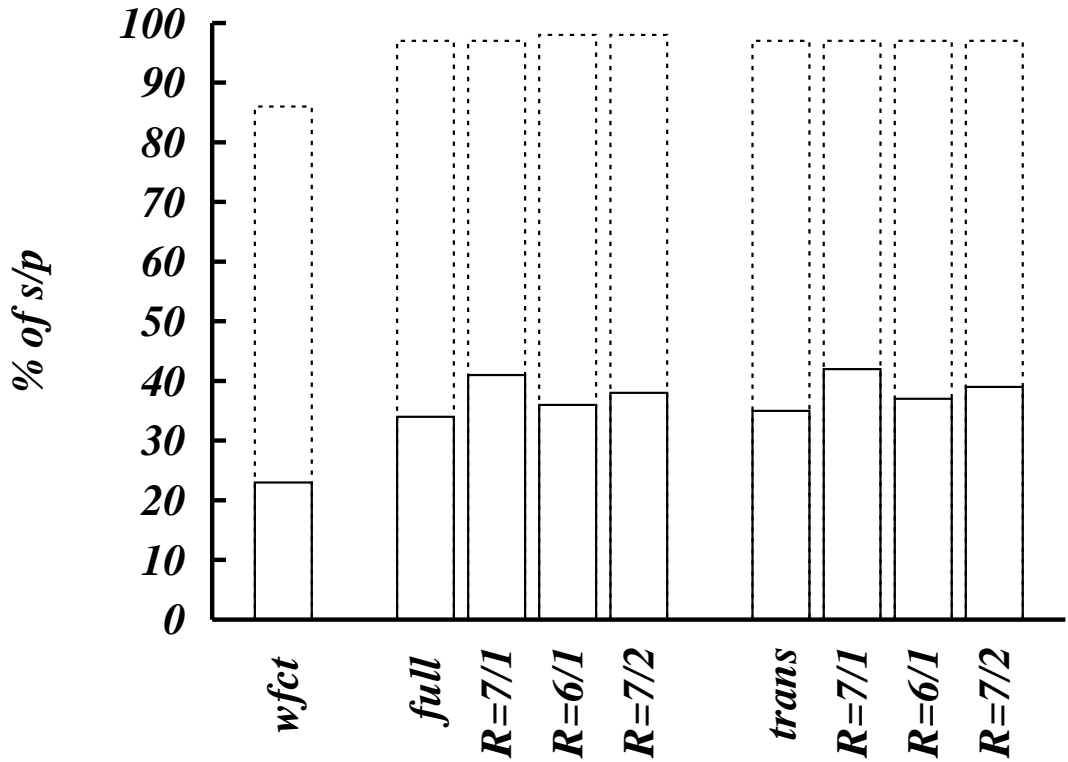}
\end{center}
\vspace*{\fill}
\caption{
Percentage of $s$- and $p$-wave in the s23 wave function
(wfct), compared to the percentages remaining in the neutron--core
final state following 1n-stripping.  The different histograms show the
results of the full calculation and the transparent limit integrated
over all impact parameters as well as at specific fixed values of
the impact parameter of the nucleon-core system (6 and 7 fm) and the
stripped neutron (1 and 2 fm).
}
\label{fig:10}
\end{figure}

\begin{figure}
\vspace*{\fill}
\begin{center}
\ForceHeight{10cm}
\BoxedEPSF{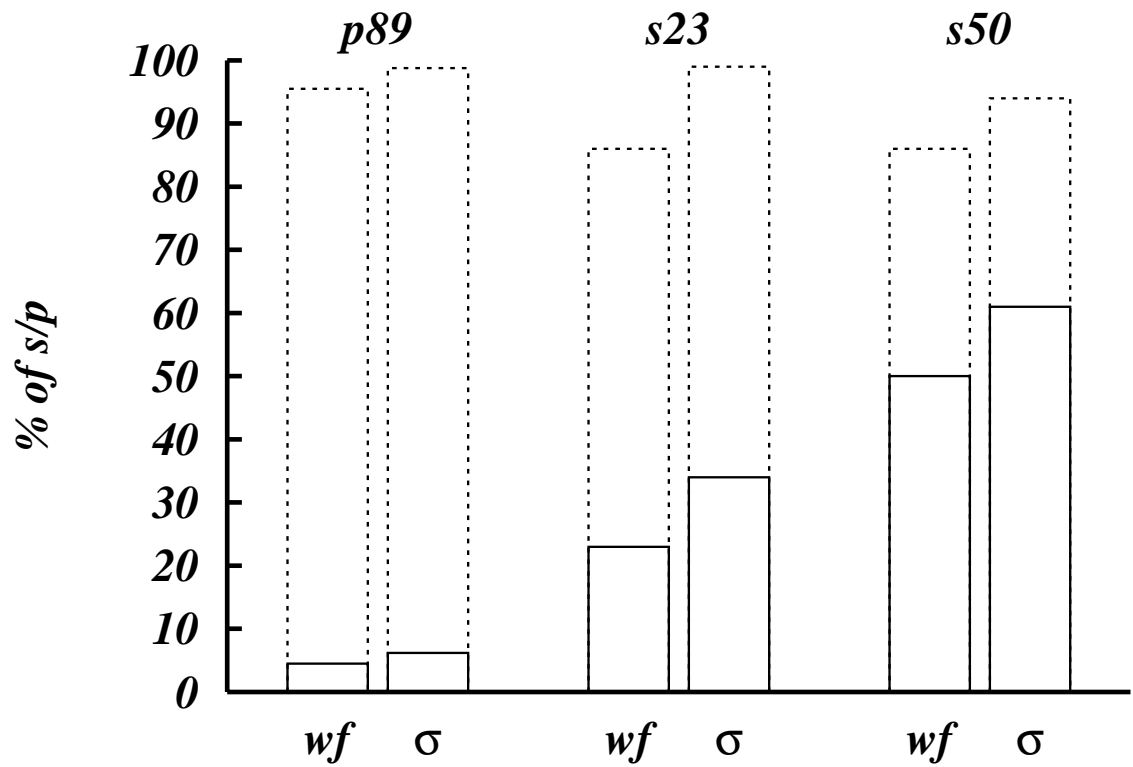}
\end{center}
\vspace*{\fill}
\caption{
Percentage of $s$- and $p$-wave in the various models,
comparing the probabilities in the wave functions (wf) to the
probabilities in the neutron--core final states ($\sigma$).
}
\label{fig:11}
\end{figure}

Next we consider the relation of the $s$- and $p$-wave probabilities
in the wave function to the corresponding probabilities in the
stripping distributions.  In principle, the stripping distributions
for $s$- and $p$-waves are sufficiently dissimilar that one should be
able to extract their relative probabilities from the data.  In
Fig.~\ref{fig:10} we show the percentage of these waves as would be found in the
energy distributions for the full calculation and also in the
no-recoil, transparent limit, using the s23 wave function.  As found
before, the transparent limit is very accurate for purposes of
extracting these ratios.  As could be anticipated from Fig.~\ref{fig:9}, it is
necessary to do a complete integration over all impact parameters to
determine the theoretical ratios.  Finally, the amount of $s$-wave in
the final state stripping distribution is systematically larger than
in the initial ground state wave function.  This is due to the larger
extension of the $s$-wave, and the resulting decreased shadowing by
the core.  In Fig.~\ref{fig:11} we compare the $s$- and $p$-wave probabilities
of the three correlated models with the final state probabilities for
1n-stripping.  In all cases the $s$-wave probability is somewhat
larger in the final state than in the wave function, for the reason
given above.

\section{The no-recoil approximation}
In this section we examine the no-recoil approximation, and will
find that it is very accurate for the \Li{} reaction.  The no-recoil
approximation is in fact exact for the integrated stripping reactions,
because the outer integration over $R_{cm}$ can be changed to an
integration over the core coordinate by a simple change of variable.
This separation cannot be made for the diffractive cross section, so
there will be some effect of including core recoil.  We discuss it
immediately below.  Following that, we discuss the energy distribution
of the single-particle stripping, which is also subject to recoil
corrections.

\subsection{Diffraction}
Diffraction arises from the fluctuations in the expectations of powers
of the profile operators (see eq. (1b)), and will only be significant at impact
parameters where the operators vary.  There are two effects that make
the diffractive scattering dependent on the distinction between the
center-of-mass coordinate and the core coordinate.  The first is that
there are no fluctuations associated with the core profile function in
the no-recoil limit, and including those fluctuations might be
expected to increase the diffractive cross section.  However, there is
another effect that goes in the opposite direction.  That is that the
spatial extension of the neutrons is smaller measured with respect to
the center of mass than with respect to the core.  One might expect a
smaller fluctuation of $S_n$ if the neutron wave functions are more
confined.  Another way to look at the question is with
eq. (\ref{sigma-difa}).  The first term in this formula can be
evaluated in either coordinate system.  The effect of the no-recoil
approximation is thus confined to the evaluation of the second term.
We have evaluated the integral numerically for the uncorrelated
$p$-wave model, and we find that the effect of recoil is to reduce the
diffractive cross section by about 20\%.  Since the corrections are
expected to scale as $1/A$, one should not ignore them for nuclei 
lighter than \Li.

\subsection{Stripping distributions}
As shown in Sec.~\ref{sec:reaction} we include the main recoil effect
(the difference in the profile functions between $R_c$ and $R_{2c}$)
in our calculation.  The only approximation is then to replace the
position $r_1$ of the stripped neutron by $-r_{2c}$,
cf. eq. (\ref{appm'}).  As this neutron is absorbed, whereas the other
neutron and the core must survive, the major contribution to the
spectrum comes from a region where $r_1$ is not small. Therefore we do
not expect to get a big effect from this. Nevertheless, we have also
calculated the exact spectrum from eq.~(\ref{mstrip}) in the
independent $p$-wave model for fixed impact parameters $R_{2c}$ and
$R_1$. The overall relative deviation of the two spectra is smaller
than 3\% with deviation near the peak being only of the order of
1/2\%. The recoil effect on the spectrum therefore seems to be
negligible.

\section{Comparison with Experiment}
\subsection{Integrated cross sections}

In Table~\ref{tab:xs} we compare the integrated cross
sections with the experimental data of ref. \cite{zi97}, Table~3.
The yields of
events having zero, one, or two neutrons in coincidence with the
$^9$Li fragment give respectively the cross sections labeled
2n-stripping, 1n-stripping, and diffraction.  
The first thing to note in the comparison with theory is that the
total two-neutron removal cross section measured, 280 mb, is much
larger than the eikonal model predicts.  There are also data on the
two-neutron removal cross section at 800 MeV/n \cite{ko89}.  Applying
the eikonal model\footnote{We note that ref. \cite{og92} obtained
241 mb, i.e., 10\% higher than experiment, in their eikonal model.}
 at this energy gives a cross section of 187~mb (s23 wave function),
only 15\% lower than the experimental value of $220\pm10$~mb.
At the lower energy the s23 model gives 179~mb.  This is close
to the theoretical value at the higher energy, which is certainly to
be expected in view of the mild change in the nucleon-nucleon cross
section between the two energies.  The direction of the change in both
the nucleon-nucleon and the eikonal removal cross section is a decrease
at the lower energy.  In contrast the experimental value is larger
at the lower energy.  The theoretical two-neutron
removal cross section behaves the same way in the case of $^6$He, 
as may be seen in the
bottom line of Table III. Here also there is fair agreement with
the experimental value at the higher energy.

Let us now turn to the individual components.  The diffractive and 
1n-stripping cross sections are within experimental error of the
most extreme wave function, s50, but the 2n-stripping cross section far
exceeds any of the models.  In principle, additional contributions to
the cross section could come from processes outside the scope of the
eikonal model.  The flux that is absorbed in the eikonal can reappear,
e.g. by multistep or rearrangement processes.  This might first show up in 
cross sections that are very small in the eikonal model--such as the
2n-stripping.  On the experimental side, the definition of the
cross sections requires some momentum cut on the neutrons to be considered as
part of the projectile fragmentation.  In ref. [6], the detector
acceptance allowed neutrons with transverse momenta of up to 60 MeV/c
to be included; the authors also quoted cross sections assuming that 20\% of
all fragmentation neutrons were outside that momentum range.  The 
deduced cross sections are given in the last column of Table III,
labeled "modified data".  

As may be seen from the table, adding the detector acceptance 
correction explains the small 2n-stripping cross section, but only
at the expense of greater disagreement with theory for the
other two cross sections.  Thus, for the diffractive cross section, we 
obtain only between 33\% and 51\% of the experimental value.  We believe 
that it not possible to explain this discrepancy within the framework of 
the eikonal model.
Recall that the unshadowed diffractive cross section has a quasibound,
$\sigma_{dif}^0 < 2 \sigma_{el}\approx 120$ mb, valid for very
extended halos, and the actually computed cross section is reduced
from this by two factors of two.  The first reduction, seen in
eq.~(\ref{dif-noshadow}), is associated with the fact that the
nucleon-target profile function blocks a significant fraction of the
halo density.  The second factor is the shadowing of the halo density
by the core-target profile function.  This shadowing is unavoidable
and reduces the diffraction cross section by an additional factor of
two.  

For the 1n stripping cross section with an assumed 20\% acceptance
correction, the eikonal model gives---
depending on the wave function--- between 63\% and 83\% of the experimental
result. The cross section increases with larger $s$-wave contribution
in the wave function. But we believe that the model that comes
closest, s50, is unrealistic on other grounds.
Until the
cross sections are better understood, it will not be possible to use
this sensitivity to test the $s$-wave probability in the wave
function.

\subsection{Relative energy spectrum}
In the analysis made in \cite{zi97} two Breit-Wigner resonances were fitted, at
$0.21\pm0.05$
and $0.62\pm0.10$~MeV, and the relative amounts of $s$- and $p$-waves in the
wave function were extracted. Let us see how this compares with our 
analysis with the calculated distributions.
As discussed in Sec.~\ref{sec:shape}, we find a peak for the final
$s$-wave at a much lower energy than the experimental spectrum shows.
However, the finite angular and energy resolution of the experimental
detectors will inevitably smear out the distribution of the extracted
relative momentum or energy, shifting the apparent peak to higher 
energy.  This should be taken into account in
comparing our spectrum with the experiments.

The authors of ref. \cite{zi97} give a table of transverse momentum
and longitudinal velocity resolution widths, and we have folded these
widths with the theoretical spectra to compare with experiment
\footnote{ Ref. [6] also quotes an
energy resolution function, ($\Delta E \sim E^{0.75}$) but this does
not go to a finite value near $E=0$, so we prefer to construct the
resolution as described.}.  The folding is done by Monte Carlo sampling.
We generate events, sampling the theoretical
distribution of $E$ in the center-of-mass system of the neutron--core
system, and assuming that the angular distribution is isotropic there.
We add a Gaussian-distributed center of mass momentum. As we do not
have any cuts in our simulation, this center of mass momentum is not
important.  We transform then from the projectile frame to the lab
frame, determining the transverse momentum as well as the Lorentz
$\beta$ in the lab frame. To these we add random errors from a
Gaussian distribution with a full width at half maximum given by
Table~2 in \cite{zi97}. The relative energy (eq. in (3.2) of
\cite{zi97}) is reconstructed using these values, which are then
binned into the same energy intervals as the experiment.

\begin{figure}
\vspace*{\fill}
\begin{center}
\ForceHeight{10cm}
\BoxedEPSF{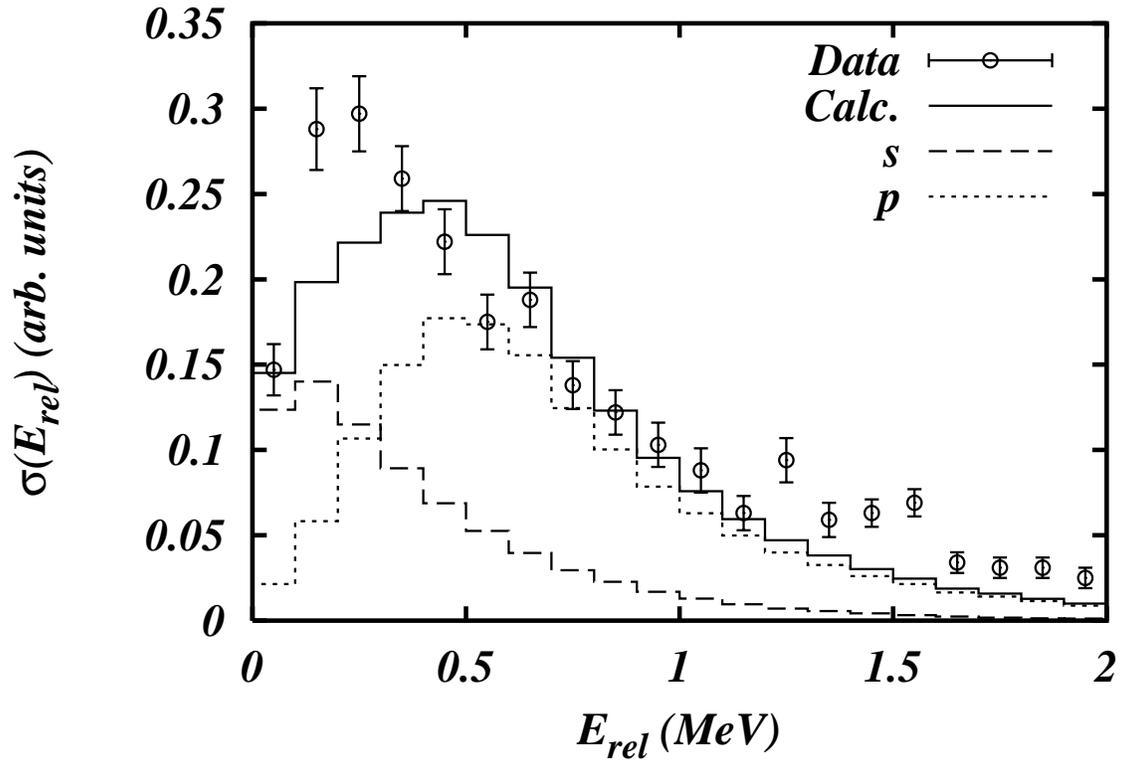}
\end{center}
\vspace*{\fill}
\caption{
Energy distribution of the s23 wave function. The histogram
is the calculated result including the experimental resolution. The
solid line is the result for all partial waves. Also shown are the
$s$- and $p$-wave components (dashed and dotted lines). The data
points are from ref. \protect\cite{zi97}.
}
\label{fig:12}
\end{figure}

Fig.~\ref{fig:12} shows the result of this procedure for the $s23$ model. The
experimental resolution increases the position of the $s$-wave peak from
0.08~MeV to 0.15~MeV, which is close to the value $0.21\pm0.05$ deduced
by the Breit-Wigner fit.  However, with our shapes for the individual
components, the measured peak at 0.21~MeV 
is a combined effect of both $s$- and $p$-wave.  The individual spectra are 
strongly overlapping, and require a realistic model of the shapes to
separate them with confidence.

In our treatment, there is no indication of a sharp $p$-wave
resonance. Our shape has a long tail
which explains most of the higher energy data. This fact can be
demonstrated by fitting the experimental data at low energies directly
to this shape.  One then finds that this curve alone is capable of
explaining almost all of the higher energy data, leaving no room 
for a resonance there.

\begin{figure}
\vspace*{\fill}
\begin{center}
\ForceHeight{10cm}
\BoxedEPSF{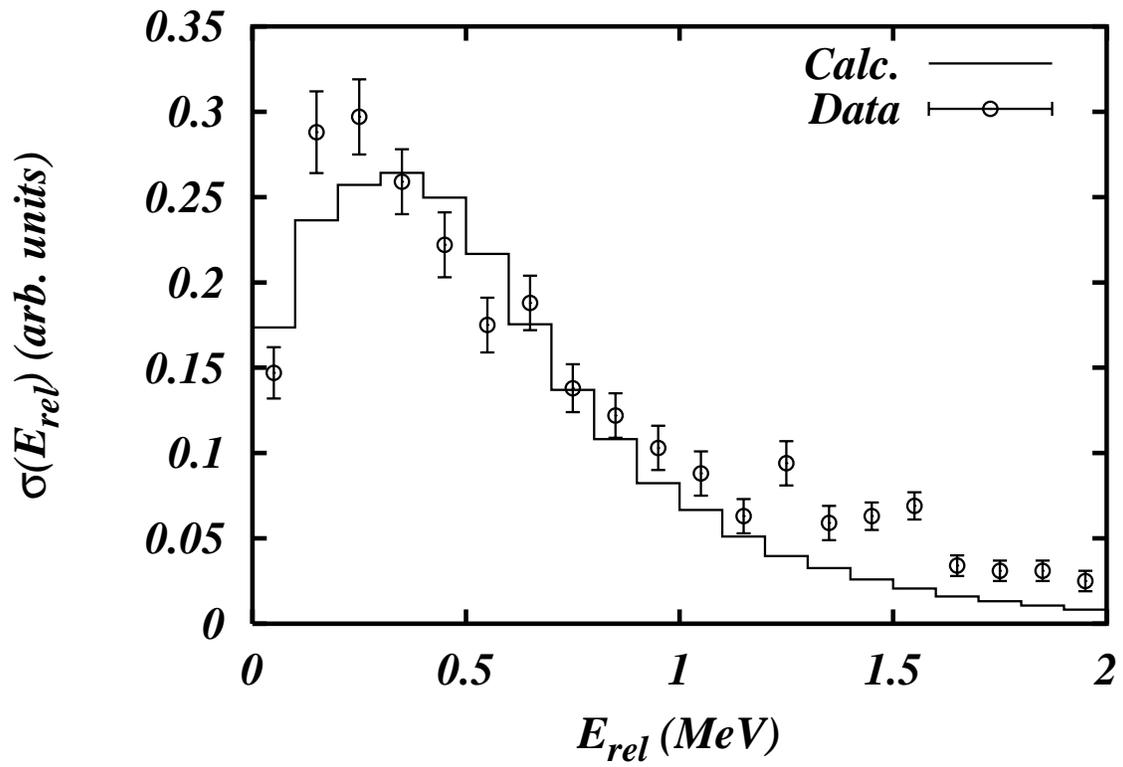}
\end{center}
\vspace*{\fill}
\caption{
``Best'' fit of the energy distribution compared to the data of
ref. \protect\cite{zi97}. The ratio of $s$- and $p$-wave and also the position
of the $p$-wave resonance was varied.
}
\label{fig:13}
\end{figure}

One of our main goals is to find out what we can conclude from this
spectrum about the $s$-wave contribution in the ground state and also
about the position of the $p$-wave resonance. We investigate this in
the following way.  The shapes of the distributions are assumed to be
the same as we found in Sec.~\ref{sec:shape}, using the parameters of 
the s23 model. We allow the ratio of 
$s$ to $p$ to vary, as well as the position of
the $p$-wave resonance (that is $E_R$ in eq. (\ref{p-analyticb})).
Using this as the input to our Monte-Carlo code we calculate
the simulated spectrum. Minimizing the $\chi^2$ with the experimental
result, we get the most likely result together with some range which
we still consider likely. Fig.~\ref{fig:13} shows the best fit to the 
experiment
we get in this way.  It is for an energy of $E_R=0.45$~MeV and an
$s$-wave component of about 40\%. From the variation of $\chi^2$ we
estimate the uncertainty in the position to be $0.1$~MeV and the
$s$-wave contribution between 30\% and 50\%. From Fig. 11 we estimate
the $s$-component in the \Li{} wave function to be
between 20\% and 40\% and the most likely value to be about 30\%.

\subsection{Transverse momentum distribution}
In order to test our final state spectrum, we also make comparisons to
a measurement of the transverse momentum distribution of neutrons.
Such a measurement has recently been performed at the same energy
\cite{zi96}.  In our analysis we assume that the momentum distribution
of the neutrons is isotropic and identical to the relative 
momentum distribution of the neutron and the core fragment.
The relative energy spectrum can then be transformed into
a transverse momentum
distribution:
$$
\frac{d\sigma}{d^2k_\perp} = \int_0^\infty dk_l \frac{1}{2\pi m k}
\frac{d\sigma}{dE}.
$$

\begin{figure}
\vspace*{\fill}
\begin{center}
\ForceHeight{10cm}
\BoxedEPSF{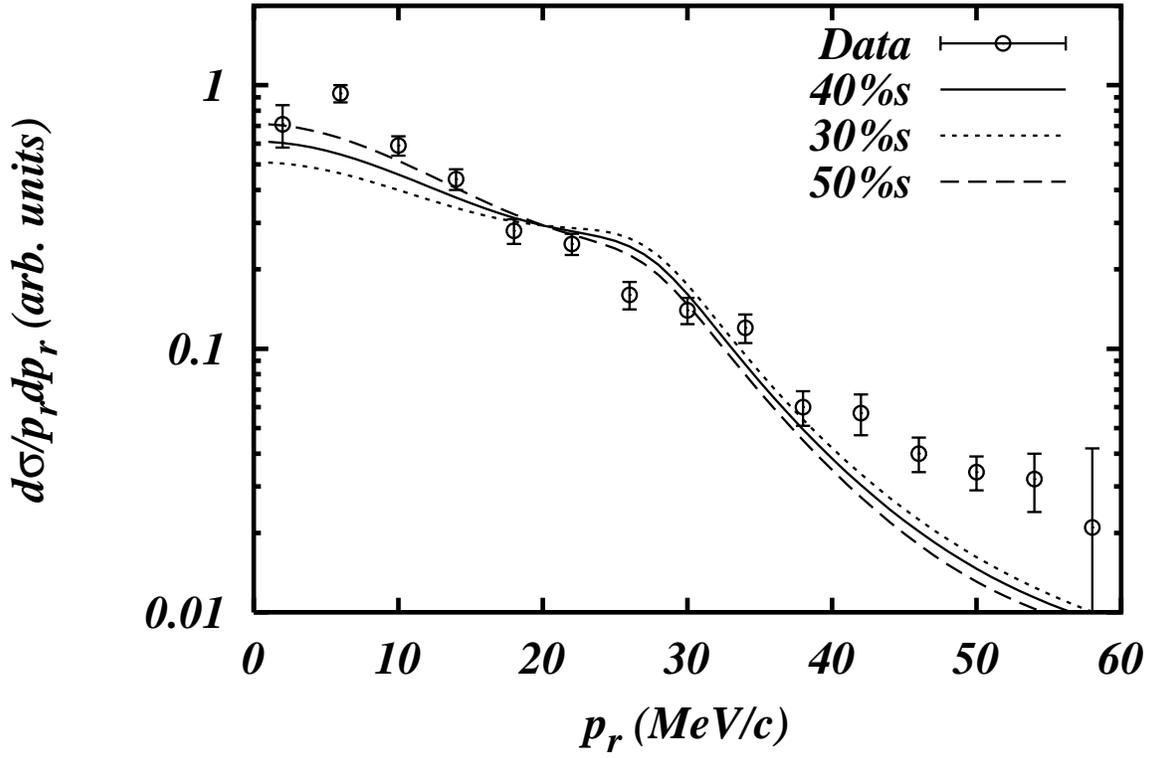}
\end{center}
\vspace*{\fill}
\caption{
Transverse momentum distribution including all partial waves
of the best fit model. Shown are the results with 30,40, and 50\%
s-wave component in the final state, corresponding to a \Li{}-ground
state component between 20 and 40 \%. The data points are from 
ref. \protect\cite{zi96}.
}
\label{fig:14}
\end{figure}

In Fig.~\ref{fig:14} we compare the result of our ``best'' model with the
experimental result. In this figure we made no attempt to 
include the experimental resolution in the theory curve.   The
agreement between theory and experiment is reasonable especially for
medium and larger momenta, whereas we have difficulties at the very
low momenta. At these momenta we expect our approximation of using the
relative momenta instead of the neutron momenta to be invalid, as
center of mass correction should play a role.  Also in this analysis
we have ignored diffraction.

\section{Conclusions and Outlook}

In this paper we applied the eikonal theory to nuclear reactions
of Borromean nuclei, with the object of developing a quantitative
tool for interpreting reaction cross sections. Because the
full theory is quite demanding from a numerical point of view, 
we examined the approximations that are commonly made.
In particular, the theory simplifies if one neglects
correlations in the ground state wave function, the final state
interaction, the distortion effects of the profile functions on unstripped
particles, or the difference between vectors referred to various
center-of-mass systems. 

The easiest cross section to interpret is the one-particle stripping,
which leaves a particle and the core in a final state of low excitation
energy.
The integrated one-particle stripping cross section can be calculated with 
rather rough approximations.  The correlations in the ground state play no
role except to determine occupation probabilities for the shell
orbitals.  In the differential cross section, the distributions of
$s$- and $p$-waves are quite distinct, allowing the occupation 
probabilities of the \Li~halo orbitals to be extracted with 
10-20\% accuracy.
Many of the simplifying approximations can be used here without
significant error.  The distortion introduced by the profile functions
have practically no effect on the shape of the distributions, and a 
moderate effect
on the extracted relative probabilities.  However it is important to include
the final state interaction in the particle-plus-core system \cite{ga96}.  
We proposed parameterizations of the $s$- and $p$-wave distributions
that take both the initial halo character and the final state
interaction into account.  The $s$-wave distribution, given by eq. 
(\ref{s-analytic}),
is derived from the asymptotic wave functions.  The $p$-wave
distribution, given by eqs. (\ref{p-analytica}-\ref{p-analyticb}), is in 
the form of a Breit-Wigner function with an energy-dependent width.  
The width varies with energy as $\Gamma\sim E^{3/2}$ at threshold, but 
the threshold region is very narrow due to the extended initial wave function.  At higher
energies the width grows more like $\Gamma\sim E^{1/2}$, which is the
characteristic behavior in the absence of a barrier.

The other two integrated cross sections are more sensitive to the
correlations in the wave function.
The two-particle stripping cross section is sensitive to the 
pairing correlation, nearly doubling when pairing is included.
If the two-nucleon stripping could be measured well, it could
be used to study this aspect of the wave function.
The other integrated cross section, the diffractive cross section,
is not only sensitive to the eikonal distortions, but we found that
it changes significantly if the recoil correlation is included in
the wave function. The differential diffractive
cross section is presently beyond our computational powers.

As a specific example, we have applied the eikonal analysis to the data
of ref. \cite{zi97}. The total two-neutron
removal cross section is larger than our calculation, and also
larger than the experimental value at a higher energy (800 MeV/n).  
This is surprising because the difference in nucleon-nucleon cross sections 
at the two energies demands the opposite trend in the eikonal model.
Our calculated one-neutron
stripping and diffractive cross sections are only 75\% and 50\%
of the measured values, respectively.  The experimental diffractive
cross section is nearly high enough to be in conflict with the bound 
given by eq. (\ref{difelb}).

In analyzing the differential cross section, we found that both 
$s$ and $p$-waves are needed.  The $p$-wave cross section does
not give a sharp resonant peak, due to the fact that the 
initial wave function is largely outside the centrifugal
barrier.  The $s$-wave distribution peaks at a very low energy,
and the experiment does not show as much of a threshold enhancement
as we predict.  While we are not able to get a perfect fit to the
data, our best fit has a $p$-wave resonance near 0.45 MeV, only
slightly lower than the measurement of
ref. \cite{yo94}. The peak
in the stripping distribution found by  \cite{zi97} is at 0.2 MeV,
and is most likely due to the combined effect of $s$- and $p$-waves.  
Again, we found that simplifying approximations of no recoil and
the transparent limit give a rather good account of the one-particle
stripping distribution.
Our analysis
with the full eikonal theory gives an $s$-wave component of between 30\% and 50\% in the
cross section, which corresponds to an $s$-wave probability in
\Li~between about 20\% and 40\%.  This range also gives a good fit
to the transverse momentum distribution of neutrons measured by ref. 
\cite{zi96}.  A number of other experimental\cite{kr93,yo94,zi96,zi97,ao96}
and theoretical studies\cite{th94,vi96} have extracted
$s$-wave probabilities of about 50\%.  An even larger value is
apparently obtained by ref. \cite{ga96}, who report 
a $p$-wave probability of 26\%.

In conclusion, the eikonal theory has considerable promise for
interpreting the distributions and cross sections in nuclear
breakup, but it has not yet proved to be a quantitative tool
for the reaction we studied.

\section{Acknowledgment}
We would like to thank W. Dostal and H. Emling for discussions
and for providing unpublished data from the experiment in ref.
\cite{zi97}.
This work was supported by the Swiss National Science Foundation
(SNF), the ``Freiwillige Akademische Gesellschaft'' (FAG) of the
University of Basel, the ``Deutsche Forschungsgemeinschaft'' (DFG),
and the U.S. Department of Energy, Nuclear Physics Division, under
Contract No. W-31-109-ENG and DE-FG-06-90ER-40561.

\begin{table}
\caption{Parameter of the different models used in comparing the
different cross sections. The first model (``p89'') uses the same
potential for $s$ and $p$ wave, whereas the other two use a deeper
potential for the $s$-wave (and all other even $L$ waves). The
strength of the $p$-wave potential is essentially fixed by the
position of the $p_{1/2}$ resonance. The lowest two entries give 
the properties of the uncorrelated pure $s$ and $p$-wave
models described in the text.}

\label{tab:models}

\begin{center}
\begin{tabular}{rcccccccc}
Model & $E_B$ & $V_{s}$ (MeV) & $a_0$ (fm) & \% $s_{1/2}$ & \% $p_{1/2}$ 
& $r^2$ (fm$^2$) & $(r_1-r_2)^2$ (fm$^2$) 
& $\left(\frac{r_1+r_2}{2}\right)^2$ (fm$^2$) \\ 
\hline
p89 &
$-0.295$ & $-35.4$ & $+1.7$ &  4.5 & 89.1 &
29.4 & 42.8 & 18.7 \\
s23&
$-0.295$ & $-47.5$ & $-5.6$ & 23.1 & 61.0 &
37.7 & 45.9 & 26.2 \\
s50 &
$-0.292$ & $-51.5$ & $-90.$ & 49.9 & 33.9 &
53.8 & 70.1 & 36.2\\
\hline
s & & & & 100 & & 45.0 & 90.0 & 22.5\\
p & & & & & 100 & 27.5 & 55.0 & 13.8
\end{tabular}
\end{center}
\end{table}

\begin{table}
\caption{Integrated cross sections (mb) with different models of the
halo nucleus \Li. }
\label{tab:xs}
\begin{center}
\begin{tabular}{lccccccc}
 &uncor. s&uncor. p& p89 & s23 & s50 &  data \cite{zi97} & modified data\\
\hline
diffraction      & 38  & 26  &  27&  33 &  40& $50\pm10$ & 77\\
1n-stripping     &  174&  123&  121& 137& 162& $170\pm20$& 195\\
1n-str. (transp.)&  182&  129&  134& 155& 182&   & \\
2n-stripping     & 4   & 3   &    6&   9&  10&  $60\pm20$& 8 \\
2n-removal       &216  & 152 & 154 & 179& 212& $280\pm30$&$280\pm30$
\\ 
\end{tabular}
\end{center}
\end{table}

\begin{table}
\caption{Integrated cross sections (mb) for {\He6} breakup on \C12.  The
experimental number is from ref. \protect\cite{ta92}.}
\label{tab:he6}
\begin{center}
\begin{tabular}{l|ccc|ccc}
 &\multicolumn{3}{c}{240~MeV/n}&\multicolumn{3}{c}{800~MeV/n}\\
 &with $S_c$&w/o $S_c$& &with $S_c$&w/o $S_c$& \\
\hline
dif      &32.3 & 64.5 & $<2\sigma_{el}=119$ &
 42.7 & 89.7 & $<2\sigma_{el}=168$ \\
1n-st  & 136 &  286 & &
 144  & 319  & \\ 
1n-st,trans & 170 &  409 & $\approx2\sigma_{re}=414$ &
 184  & 488  & $\approx2\sigma_{re}=493$\\
2n-st    &16.7 & 61.6 & &
 19.8 & 84.1 & \\
\hline
$\sigma_{-2n}$   & 185 & & &
 206  &  & $\sigma_{exp}=189\pm14$
\end{tabular}
\end{center}
\end{table}

\newpage
\newpage
\newpage
\newpage
\newpage
\newpage
\newpage
\newpage
\newpage
\newpage
\newpage
\newpage
\newpage
\end{document}